\documentclass[10pt,showpacs,floatfix,letterpaper,amsmath,amsfonts,amssymb,aps,prl,twocolumn]{revtex4-1}
\usepackage[latin1]{inputenc}
\usepackage{bm}
\usepackage{graphicx}
\usepackage{tensor}

\newcommand{\op}[1]{\hat{#1}}
\newcommand{\ket}[1]{\lvert #1\rangle}
\newcommand{\bra}[1]{\langle #1 \rvert}
\newcommand{\pr}[1]{\ket{#1}\bra{#1}}
\newcommand{\ipr}[2]{\langle #1 \vert #2 \rangle}
\newcommand{\mean}[1]{\left\langle #1 \right\rangle}
\newcommand{\cmean}[2]{\tensor[_{#1}]{\mean{#2}}{}}

\newcommand{\Tr}[1]{\text{Tr}[#1]}
\newcommand{\Trs}[1]{\text{Tr}_S[#1]}
\newcommand{\Trd}[1]{\text{Tr}_D[#1]}

\begin{document}
\title{Weak values are universal in von Neumann measurements}
\author{Justin Dressel}
\author{Andrew N. Jordan}
\affiliation{Department of Physics and Astronomy, University of Rochester, Rochester, New York 14627, USA}

\date{\today}

\begin{abstract}
We refute the widely held belief that the quantum weak value necessarily pertains to weak measurements.  To accomplish this, we use the transverse position of a beam as the detector for the conditioned von Neumann measurement of a system observable.  For any coupling strength, any initial states, and any choice of conditioning, the averages of the detector position and momentum are completely described by the real parts of three generalized weak values in the joint Hilbert space.  Higher-order detector moments also have similar weak value expansions.  Using the Wigner distribution of the initial detector state, we find compact expressions for these weak values within the reduced system Hilbert space.  As an application of the approach, we show that for any Hermite-Gauss mode of a paraxial beam-like detector these expressions reduce to the real and imaginary parts of a single system weak value plus an additional weak-value-like contribution that only affects the momentum shift.
\end{abstract}

\pacs{03.65.Ta,03.65.Ca,03.67.-a}
\maketitle

Since its introduction in 1988 by Aharonov, Albert, and Vaidman (AAV) \cite{Aharonov1988,*Duck1989} and subsequent confirmation \cite{Ritchie1991,Pryde2005}, the weak value of a quantum observable has been a source of considerable controversy.  AAV showed that a \emph{weak} conditioned von Neumann measurement which coupled an observable $\op{A}$ to a continuous detector consistently produced the complex weak value expression, $\mean{A}^w = \bra{\psi_f}\op{A}\ket{\psi_i}/\ipr{\psi_f}{\psi_i}$ in the detector's linear response after pre-selecting the system state to $\ket{\psi_i}$ and post-selecting the system state to $\ket{\psi_f}$.  Notably, the parts of this complex expression need not be constrained to the eigenvalue range of $\op{A}$, a fact which has prompted considerable recent interest both for amplifying the measurements of small quantities in weak measurements \cite{Hosten2008,Dixon2009,*Starling2010,*Starling2010b} and for fruitfully using weak measurements to interpret quantum phenomena \cite{Steinberg1995,Resch2004,Lundeen2009,Kocsis2011,Williams2008,*Goggin2011,Dressel2011,Hofmann2011}.  

There has also been considerable recent interest in generalizing the derivation of pre- and post-selected measurements beyond the weak measurement regime considered by AAV.  Example efforts include the increase of the coupling strength \cite{Williams2008,Dressel2011,Geszti2010,Haapasalo2011}, the addition of detector dynamics \cite{Jozsa2007,DiLorenzo2008}, the addition of decoherence and noise \cite{Shikano2010,*Shikano2011}, treatments of orthogonal post-selections \cite{Wu2011}, considerations of full counting statistics \cite{DiLorenzo2012}, a realization with Fock states \cite{Simon2011}, and the determination of optimal detector states \cite{Shikano2012}.  The AAV regime weak value has also been generalized to mixed initial states $\op{\rho}_i$ and arbitrary post-selections represented by positive operators $\op{P}_f$ \cite{Wiseman2002,Dressel2010,*Dressel2012,*Dressel2012b,Dressel2012d},
\begin{align}\label{eq:wv}
  \mean{A}^w = \frac{\Tr{\op{P}_f \, \op{A} \, \op{\rho}_i}}{\Tr{\op{P}_f \, \op{\rho}_i}}.
\end{align}
Notably, Eq.~\eqref{eq:wv} reduces to the original expression when $\op{\rho}_i = \pr{\psi_i}$ and $\op{P}_f = \pr{\psi_f}$, but also has the benefit of subsuming the expectation value of $\op{A}$ as a special case when $\op{P}_f = \op{1}$.

In this Letter, we extend these works with five main results.  Our primary result is to show that all von Neumann measurements are exactly described by generalized weak values such as Eq.~\eqref{eq:wv} for any coupling strength, any choice of initial mixed system or detector states, and any choice of generalized post-selection.  Hence, \emph{weak values are universal in von Neumann measurements}, and thus are not solely peculiarities of the AAV weak measurement regime.  Our second and third results are compact expressions for the relevant generalized weak values in terms of the Wigner distribution of the detector.  Finally, our fourth and fifth results are applications of our general results to transverse Hermite-Gaussian modes of a detecting beam, such as those naturally produced by laser cavities.  In the Supplementary Material \footnote{See Supplementary Material for generalizations to higher-order detector moments, as well as arbitrary Hermite-Gauss mode superpositions for the detector.} we further generalize our main results to higher-order detector moments and arbitrary Hermite-Gauss detector superpositions for completeness.

\emph{Conditioned von Neumann measurement}.---
Consider a von Neumann measurement \cite{Aharonov1988}, which consists of an impulsive interaction Hamiltonian of the form $\op{H}_I = g \delta(t-t_0) \op{A}\otimes\op{p}$, where $\op{A}$ is an observable on the \emph{system} Hilbert space that we wish to measure and $\op{p}$ is the transverse momentum on a \emph{detector} Hilbert space.  Solving the Schr\"{o}dinger equation $i\hbar\partial_t \op{U} = \op{H}_I \op{U}$ with this interaction produces the unitary evolution operator $\op{U}_g = \exp( g\op{A}\otimes\op{p} / i\hbar )$, which generates translations in $\op{x}$ by an amount $g \op{A}$ due to the canonical commutation relations $[\op{x},\op{p}] = i\hbar$.

Now consider the following experimental procedure.  First, prepare an arbitrary joint state of the system and detector, represented by a density operator $\op{\rho}_{SD}$.  Second, apply the impulsive interaction $\op{U}_g$.  Third, measure the detector position $\op{x}$ or momentum $\op{p}$.  Finally, condition the detector measurements on an arbitrary generalized post-selection on the system, which can always be represented by a positive probability operator $\op{P}_f$ \cite{Dressel2012b,Dressel2012d}.  

The conditioned detector averages measured in the laboratory will then have the exact form \cite{Dressel2012d},
\begin{subequations}\label{eq:detresraw}
\begin{align}
  \cmean{f}{x} &= \frac{\Tr{(\op{P}_f\otimes\op{x})\, \op{\rho}_{SD}' }}{\Tr{(\op{P}_f\otimes\op{1}_D)\, \op{\rho}_{SD}'}}, \displaybreak[0]\\
  \cmean{f}{p} &= \frac{\Tr{(\op{P}_f\otimes\op{p})\, \op{\rho}_{SD}' }}{\Tr{(\op{P}_f\otimes\op{1}_D)\, \op{\rho}_{SD}'}},
\end{align}
\end{subequations}
where $\op{\rho}_{SD}' = \op{U}_g \op{\rho}_{SD} \op{U}_g^\dagger$ is the entangled joint post-interaction state at a time $t>t_0$.

As written, Eqs.~\eqref{eq:detresraw} show that the joint observables $\op{P}_f\otimes\op{x}$ and $\op{P}_f\otimes\op{p}$ are averaged with respect to the final joint state $\op{\rho}_{SD}'$.  However, we can also express these averages in terms of the \emph{initial} joint state by commuting the detector observables symmetrically past the evolution operators $\op{U}_g$ to obtain our primary result,
\begin{subequations}\label{eq:cav}
\begin{align}
  \cmean{f}{x} &= \text{Re}\mean{x}^w + g\, \text{Re}\mean{A}^w, \\
  \cmean{f}{p} &= \text{Re}\mean{p}^w. 
\end{align}
\end{subequations}

\begin{figure}[t]
  \begin{center}
  \includegraphics[width=0.2\columnwidth]{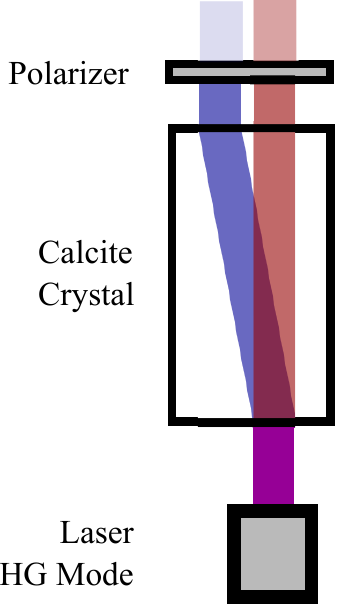}\quad
  \includegraphics[width=0.6\columnwidth]{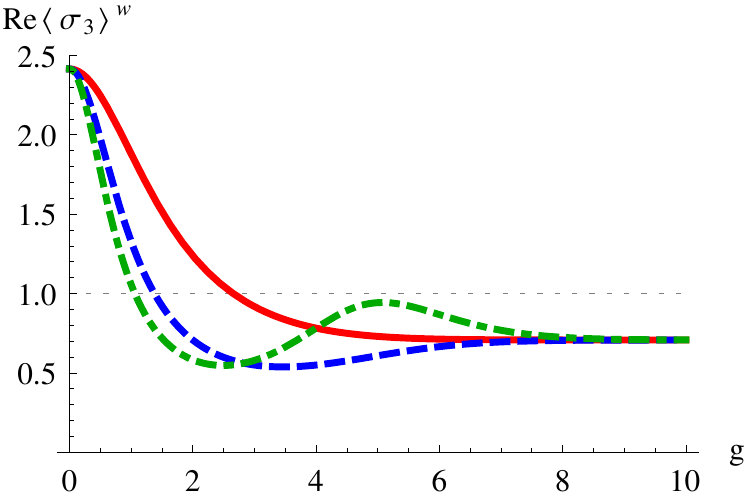}
  \end{center}
  \caption{(color online) (left) A possible implementation of a conditioned polarization measurement similar to \cite{Ritchie1991}, where the length of a birefringent crystal determines the coupling strength $g$.  (right) The weak value $\text{Re}\mean{\sigma_3}^w$ corresponding to the Hermite-Gauss detector profiles in Fig.~\ref{fig:detsig} with $m=0$ (solid, red), $m=1$ (dashed, blue), and $m=2$ (dot-dashed, green), obtained by averaging according to Eq.~\eqref{eq:cavhg}.  The weak limit $g\to 0$ is identical for all detectors, as is the strong limit $g\to\infty$ of a classical conditioned average, but the specifics of the transition depend on how the detector decoheres the state.  The dotted horizontal line is the eigenvalue bound of $1$.}
  \label{fig:wv}
\end{figure}

The averages are exactly characterized by the real parts of three \emph{generalized weak values} \cite{Wiseman2002,Dressel2010,Dressel2012,Dressel2012d,Dressel2012b} that are of the form \eqref{eq:wv}, but are on the joint Hilbert space of the system and detector,
\begin{subequations}\label{eq:wvs}
\begin{align}
  \label{eq:awv}
  \mean{A}^w &= \frac{\Tr{\op{P}_{SD}'\, (\op{A}\otimes\op{1}_D)\,\op{\rho}_{SD}}}{\Tr{\op{P}_{SD}'\,\op{\rho}_{SD}}}, \displaybreak[0]\\
  \label{eq:xwv}
  \mean{x}^w &= \frac{\Tr{\op{P}_{SD}'\, (\op{1}_S\otimes\op{x})\,\op{\rho}_{SD}}}{\Tr{\op{P}_{SD}'\,\op{\rho}_{SD}}}, \\
  \label{eq:pwv}
  \mean{p}^w &= \frac{\Tr{\op{P}_{SD}'\, (\op{1}_S\otimes\op{p})\,\op{\rho}_{SD}}}{\Tr{\op{P}_{SD}'\,\op{\rho}_{SD}}}. 
\end{align}
\end{subequations}
The pre-selection for each weak value is equal to the initial joint state $\op{\rho}_{SD}$, while the post-selection is equal to the Heisenberg-evolved joint post-selection operator, $\op{P}_{SD}' = \op{U}^\dagger_g (\op{P}_f\otimes\op{1}_D) \op{U}_g$.  As noted before, when $\op{P}_f = \op{1}_S$ there is no post-selection and the weak values \eqref{eq:wvs} will reduce to expectation values as a special case.  The higher-order detector moments are provided in the Supplementary Material \footnotemark[\value{footnote}], and all have similar expansions into joint weak values.

Importantly, these relations hold for any coupling strength $g$, any (possibly entangled) initial joint state $\op{\rho}_{SD}$, and any generalized post-selection $\op{P}_f$; that is, \emph{all von Neumann detector (conditioned) averages are exactly described by generalized weak values.}  This important result seems to have been missed in the existing literature due to the fact that the generalized weak values \eqref{eq:wvs} cannot be written in a form with projective pre- and post-selections as defined originally by AAV \cite{Aharonov1988}.  Moreover, they explicitly include the detector information, so are not solely system quantities.

\begin{figure}[t]
  \begin{center}
  \includegraphics[width=0.9\columnwidth]{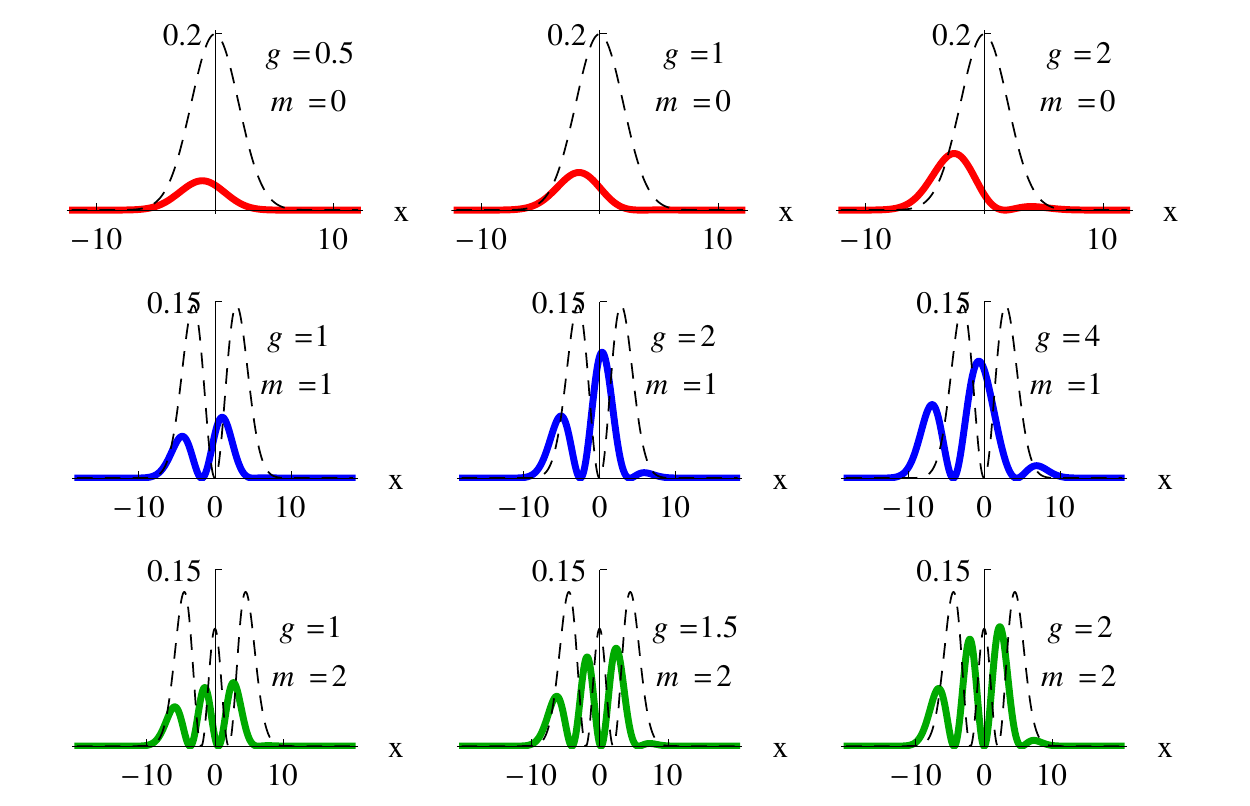}
  \end{center}
  \caption{(color online) Post-selected detector intensities for the initial polarization state $\ket{\psi_i} = (\cos(7\pi/8),\sin(7\pi/8))$ and final post-selection $\ket{\psi_f} = (1,1)/\sqrt{2}$, using the first three Hermite-Gauss detector modes with $\sigma = 2$.  Averaging these profiles produces weak values according to Eq.~\eqref{eq:cavhg} and shown in Fig.~\ref{fig:wv}.  The dashed line indicates the initial detector intensity.}
  \label{fig:detsig}
\end{figure}

\emph{Reduced state expressions}.---
If we prepare a product initial state $\op{\rho}_{SD} = \op{\rho}_S\otimes\op{\rho}_D$, where $\op{\rho}_S$ ($\op{\rho}_D$) is the initial state of the system (detector), then we can exploit the product form of the observables to further simplify Eqs.~\eqref{eq:wvs}.  Notably, since $[\op{A},\op{U}_g]=0$ we can express Eq.~\eqref{eq:awv} as a weak value only on the system Hilbert space,
\begin{align}\label{eq:awvs}
  \mean{A}^w &= \frac{\Trs{\op{P}_f\, \op{A}\, \op{\rho}_S'}}{\Trs{\op{P}_f\, \op{\rho}_S'}},
\end{align}
where the pre-selection state $\op{\rho}_S'$ is the reduced system state \emph{after} the interaction, $\op{\rho}_S' = \Trd{\op{\rho}_{SD}'}$, and $\Trs{\cdot}$ ($\Trd{\cdot}$) is the partial trace over the system (detector) Hilbert space.  All detector information has been absorbed into an effective preparation of the reduced system state $\op{\rho}_S'$.

Since the joint post-interaction state $\op{\rho}_{SD}'$ is necessarily entangled by the interaction, the reduced system state $\op{\rho}_S'$ in \eqref{eq:awvs} will be mixed.  However, for sufficiently weak coupling one can approximately neglect the interaction in \eqref{eq:awvs} and substitute the initial system state $\op{\rho}_S' \to \op{\rho}_S$.  The detector response \eqref{eq:cav} will then be linear in $g$ and match the original observation of AAV \cite{Aharonov1988} as an approximate special case.

By introducing the Wigner distribution of the detector state $W_D(x,p) = \frac{1}{2\pi\hbar}\int dy\, \bra{x-y/2}\op{\rho}_D\ket{x+y/2}\, e^{i p y/\hbar}$ and its Fourier transform $\widetilde{W}_D(x,y) = \int dp\, W_D(x,p)\, e^{-i p y/\hbar} = \bra{x-y/2}\op{\rho}_D\ket{x+y/2}$, we can express the exact reduced system state $\op{\rho}_S'$ in a useful and compact form, which is our second main result,
\begin{align}
  \label{eq:sysred}
  \op{\rho}_S' &= \int dx \, \widetilde{W}_D(x,g\,\text{ad}[\op{A}])(\op{\rho}_S).
\end{align}
Here $\text{ad}[\op{A}](\op{B}) = \op{A}\op{B}-\op{B}\op{A}$ is the adjoint left-action of $\op{A}$ as a commutator operation.

\begin{figure}[t]
  \begin{center}
  \includegraphics[width=0.8\columnwidth]{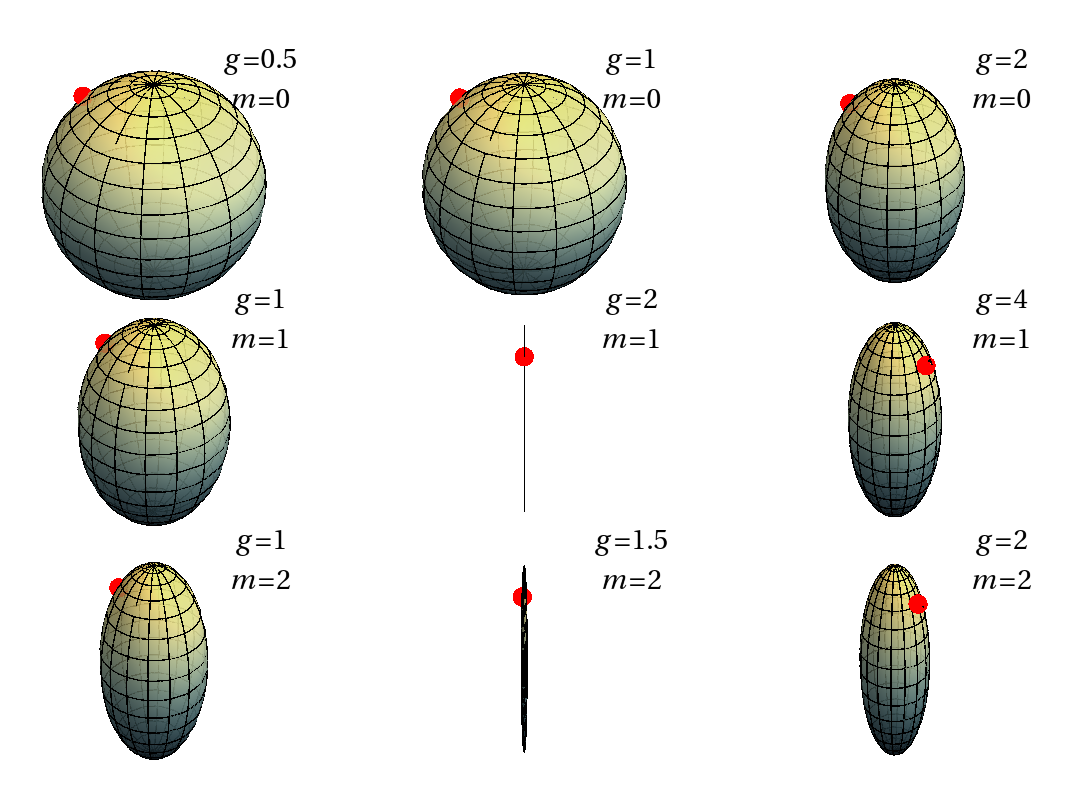}
  \end{center}
  \caption{(color online) Reduced polarization states corresponding to the detector responses in Fig.~\ref{fig:detsig}. Given an initial state $\op{\rho}_S = (\op{1} + \sum_i r_i \op{\sigma}_i)/2$ with Pauli operators $\op{\sigma}_i$ and measurement of $\op{A} = \op{\sigma}_3$ with Hermite-Gauss mode $m$, the post-interaction state from Eq.~\eqref{eq:sysredhg} is $\op{\rho}_{S,m}' = (\op{1} + r_3 \op{\sigma}_3 + L_m[(g/\sigma)^2] \exp(-(g/\sigma)^2/2)(r_1 \op{\sigma}_1 + r_2 \op{\sigma}_2))/2$.  Bloch sphere distortions are shown with the $\sigma_3$ axis aligned vertically; the red dot tracks the initial state chosen in Fig.~\ref{fig:detsig}.  For $m>0$ any initial state will experience decoherence oscillations and pass directly through the $\sigma_3$ axis before partially recohering.}
  \label{fig:bloch}
\end{figure}

To directly compare the joint weak values Eqs.~\eqref{eq:xwv} and \eqref{eq:pwv} with \eqref{eq:awvs}, we also express them within the system Hilbert space,
\begin{subequations}\label{eq:xpwvsys}
\begin{align}
  \text{Re}\mean{x}^w &= \frac{\Trs{\op{P}_f\, \mathcal{X}(\op{\rho}_S)}}{\Trs{\op{P}_f \op{\rho}_S'}}, \\
  \text{Re}\mean{p}^w &= \frac{\Trs{\op{P}_f\, \mathcal{P}(\op{\rho}_S)}}{\Trs{\op{P}_f \op{\rho}_S'}},
\end{align}
\end{subequations}
by introducing the operations $\mathcal{X}(\op{\rho}_S) = \Trd{\op{U}_g (\op{\rho}_S\otimes(\op{x}\op{\rho}_D + \op{\rho}_D \op{x})/2) \op{U}_g^\dagger}$ and $\mathcal{P}(\op{\rho}_S) = \Trd{\op{U}_g (\op{\rho}_S\otimes(\op{p}\op{\rho}_D + \op{\rho}_D \op{p})/2) \op{U}_g^\dagger}$ that act upon the initial system state.  The Weyl-ordered operator products that appear allow us to use the Fourier transformed Wigner distribution of the detector in \eqref{eq:sysred} to find compact expressions for these operations, which is our third main result,
\begin{subequations}\label{eq:ops}
\begin{align}
  \label{eq:xop}
  \mathcal{X}(\op{\rho}_S) &= \int dx \, x \, \widetilde{W}_D(x,g\, \text{ad}[\op{A}])(\op{\rho}_S), \displaybreak[0]\\
  \label{eq:pop}
  \mathcal{P}(\op{\rho}_S) &= i\hbar \left[\partial_z\int dx\, \widetilde{W}_D(x,z)\right]_{z\to g\,\text{ad}[\op{A}]}(\op{\rho}_S), \\
  &= i\hbar\, \partial_{g\,\text{ad}[\op{A}]}\, \op{\rho}_S'. \nonumber
\end{align}
\end{subequations}
Notably, Eq.~\eqref{eq:pop} allows one to simply obtain the momentum response once the functional form of reduced system state \eqref{eq:sysred} is known. Generalizations to higher-order detector moments are provided in the Supplementary Material \footnotemark[\value{footnote}].

\emph{Hermite-Gauss modes}.---
To show how our general results in Eqs.~\eqref{eq:cav}, \eqref{eq:awvs}, \eqref{eq:sysred}, \eqref{eq:xpwvsys}, and \eqref{eq:ops} can be applied, we now consider the Hermite-Gauss modes $\{\ket{h_m}\}$, which are a widely used complete set of transverse modes naturally generated in laser cavities that can describe an initial zero-mean and collimated detecting beam.  The Wigner distribution for a Hermite-Gauss mode of order $m\in\{0,1,2,\dots\}$ has the form \cite{Simon2000},
\begin{subequations}\label{eq:hgwigner}
\begin{align}
  W^{\text{HG}}_m(x,p) &= \frac{(-1)^m}{\pi \hbar}\, L_m[2 G(x,p)]\, e^{-G(x,p)}, \\
  G(x,p) &= \frac{x^2}{2\sigma^2} + \frac{2\sigma^2p^2}{\hbar^2},
\end{align}
\end{subequations}
where $L_m$ is a Laguerre polynomial of order $m$.  The first few such polynomials are shown in Table~\ref{tab:poly} for reference.  

\begin{table}
  \begin{tabular}{c c c}
    m & $\text{L}_m(x)$ & $-2\,\text{L}'_m(x)$ \\ \hline
    0 & 1 & 0 \\
    1 & $1-x$ & 2 \\
    2 & $1-2x+x^2/2$ & $4-2x$ \\
    3 & \; $1-3x+3x^2/2-x^3/6$ \; & $6-6x+x^2$
  \end{tabular}
  \caption{Laguerre polynomials $\text{L}_m(x)$ and their derivatives for the first few $m$.  These polynomials appear naturally for Hermite-Gauss modes in their Wigner distribution \eqref{eq:hgwigner}, as well as the resulting system operations \eqref{eq:sysredhg} and \eqref{eq:delta}.}
  \label{tab:poly}
\end{table}

After Fourier-transforming Eq.~\eqref{eq:hgwigner} and integrating according to Eq.~\eqref{eq:sysred}, we obtain a compact expression for the exact post-interaction reduced system state for any coupling strength and initial detector mode $m$, which is our fourth main result,
\begin{align}\label{eq:sysredhg}
  \op{\rho}_{S,m}' &= L_m\left[ -2\,\epsilon\, \mathcal{L}[\op{A}] \right]\,e^{\epsilon\,\mathcal{L}[\op{A}]}(\op{\rho}_S).
\end{align}
Notably, a measurement strength parameter $\epsilon = (g/2\sigma)^2$ naturally appears for all modes along with the \emph{Lindblad} operation $\mathcal{L}[\op{A}] = -\text{ad}^2[\op{A}]/2$ that decoheres bases orthogonal to the eigenbasis of $\op{A}$ \cite{Breuer2007,Dressel2012d}.  Furthermore, the functional form of \eqref{eq:sysredhg} is the same as the Wigner distribution \eqref{eq:hgwigner} up to normalization, but with the function $G(x,p)$ replaced by the Lindblad operation $-\epsilon \mathcal{L}[\op{A}]$.  Superpositions of modes are considered in the Supplementary Material \footnotemark[\value{footnote}].

Using Eqs.~\eqref{eq:sysredhg}, \eqref{eq:ops}, and \eqref{eq:cav} we obtain the following compact results for the exact detector averages for any initial Hermite-Gauss detector mode of order $m$, which is our fifth and final main result,
\begin{subequations}\label{eq:cavhg}
\begin{align}
  \cmean{f}{x} &= g\, \text{Re}\mean{A}^w, \\
  \label{eq:cavhgp}
  \cmean{f}{p} &= g\, \frac{\hbar}{(2\sigma)^2}\,2\, \text{Im}(\mean{A}^w + \Delta_m).
\end{align}
\end{subequations}
Perhaps surprisingly, they are completely parametrized by a single generalized system weak value \eqref{eq:awvs} with pre-selection equal to the reduced post-interaction system state $\op{\rho}_{S,m}'$ given in Eq.~\eqref{eq:sysredhg}, and one additional weak-value-like correction term for the higher mode numbers $m\ge 1$,
\begin{subequations}\label{eq:delta}
\begin{align}
  \Delta_m &= \frac{\Trs{\op{P}_f\, \op{A}\, \mathcal{M}_m(\op{\rho}_S)}}{\Trs{\op{P}_f\, \op{\rho}_{S,m}'}}, \\
  \mathcal{M}_m(\op{\rho}_S) &= -2\, L'_m\left[ -2\, \epsilon\,\mathcal{L}[\op{A}] \right]\,e^{\epsilon\,\mathcal{L}[\op{A}]}(\op{\rho}_S).
\end{align}
\end{subequations}
The first few polynomials $-2\,L'_m(x)$ in $\mathcal{M}_m$ that contain the derivatives of Laguerre polynomials are shown in Table~\ref{tab:poly} for reference.

The appearance of a correction to $\text{Im}\mean{A}^w$ in Eq.~\eqref{eq:cavhgp} further strengthens the observation in \cite{Hofmann2011,Dressel2012d} that $\text{Im}\mean{A}^w$ pertains solely to the rate of change of the post-selection probability and not to the measurement of $\op{A}$ itself.  Indeed, for $m=0$ Eqs.~\eqref{eq:sysredhg} and \eqref{eq:cavhg} correctly reproduce the exact Gaussian detector case that we derived in more detail using a different method in \cite{Dressel2012d}.  

We stress that these are general results for any system observable $\op{A}$.  Figs.~\ref{fig:wv}, \ref{fig:detsig}, and \ref{fig:bloch} show the special case of an optical application, where $\op{A} = \op{\sigma}_3$ is a polarization observable being measured by a Hermite-Gaussian beam.  Fig.~\ref{fig:detsig} shows post-interaction spatial intensity profiles for the detector, while Fig.~\ref{fig:bloch} shows the corresponding reduced polarization states.  Fig.~\ref{fig:wv} shows a possible implementation of this example that is analogous to the experiment performed in \cite{Ritchie1991}, as well as how the generalized weak value \eqref{eq:awvs} continuously changes into a classical conditioned average as the initial state decoheres.

\emph{Conclusions}.---
Throughout the controversial history of the quantum weak value \eqref{eq:wv}, it has been tacitly assumed that it was a peculiarity specific to the AAV weak measurement regime.  We have shown in this letter that such an assumption has been unwarranted.  Indeed, we have shown that \emph{all} (conditioned) averages for any von Neumann detector \eqref{eq:cav} will be completely characterized by three generalized weak values \eqref{eq:wvs} on the joint Hilbert space of the system and detector, which makes such weak values a universal feature of von Neumann measurements.

We have also shown how to obtain practical and compact operational expressions for these weak values on the system space alone in terms of the reduced post-interaction system state \eqref{eq:sysred} and two additional operations \eqref{eq:ops}.  In the process, we have highlighted the pragmatic importance of the Fourier transformed Wigner distribution of the detector for describing how the detector decoheres the system due to the interaction.

Finally, we have shown that for arbitrary Hermite-Gauss modes of a beam-like detector, we obtain simple and intuitive operational expressions for the reduced system state \eqref{eq:sysredhg} and the (conditioned) detector averages \eqref{eq:cavhg} that involve the Lindblad decoherence operation.  The detector averages contain only the real and imaginary parts of a single system weak value \eqref{eq:awvs}, along with a correction \eqref{eq:delta} to the imaginary part that appears only for the momentum average with higher-order modes.  

Not all observable measurements use such a von Neumann detector, and not all von Neumann detectors operate impulsively on the time scales of the system or the detector.  However, a sufficiently wide class of observable measurements use such an impulsive von Neumann procedure that the original weak value paper \cite{Aharonov1988} dubbed it the ``standard measuring procedure,''  so its universal description with generalized weak values is important.  We also stress that the generalized weak value appears under reasonable conditions even in the general treatment of observable measurements that we developed in detail in \cite{Dressel2010}, a fact which warrants further scrutiny in light of the universality shown here.

We acknowledge support from the National Science Foundation under Grant No. DMR-0844899, and the US Army Research Office under grant Grant No. W911NF-09-0-01417.

%
\clearpage
\pagebreak[0]
\appendix
\section*{Supplementary Material}
\subsection*{Detector Moments}
All detector moments can be determined through conditioned characteristic functions,
\begin{subequations}
\begin{align}
  \cmean{f}{e^{i \lambda x}} &= \frac{\Tr{(\op{P}_f\otimes e^{i \lambda \op{x}})\,\op{U}_g\op{\rho}_{SD}\op{U}^\dagger_g}}{\Tr{(\op{P}_f\otimes \op{1}_D)\,\op{U}_g\op{\rho}_{SD}\op{U}^\dagger_g}}, \\
  &= \frac{\Trs{\op{P}_f\, e^{i \lambda g \op{A}}\, \mathcal{X}_\lambda(\op{\rho}_{SD})}}{\Trs{\op{P}_f\, \op{\rho}'_S}}, \nonumber \\
  \cmean{f}{e^{i \lambda p}} &= \frac{\Tr{(\op{P}_f\otimes e^{i \lambda \op{p}})\,\op{U}_g\op{\rho}_{SD}\op{U}^\dagger_g}}{\Tr{(\op{P}_f\otimes \op{1}_D)\,\op{U}_g\op{\rho}_{SD}\op{U}^\dagger_g}}, \\
  &= \frac{\Trs{\op{P}_f\, \mathcal{P}_\lambda(\op{\rho}_{SD})}}{\Trs{\op{P}_f\, \op{\rho}'_S}},\nonumber  
\end{align}
\end{subequations}
where we have used the Weyl relation \cite{Alicki2001}, $e^{i a \op{x}} e^{-i b \op{p}/\hbar} = e^{i a b} e^{-i b \op{p}/\hbar} e^{i a \op{x}}$, and have defined the post-interaction reduced state $\op{\rho}'_S = \Trd{\op{U}_g \op{\rho}_{SD} \op{U}^\dagger_g}$, as well as the $\lambda$-dependent operations,
\begin{subequations}\label{eq:opslambda}
\begin{align}
  \mathcal{X}_\lambda(\op{\rho}_{SD}) &= \frac{1}{2}\Trd{\op{U}_g (e^{i\lambda \op{x}} \op{\rho}_{SD} + \op{\rho}_{SD}e^{i\lambda \op{x}}) \op{U}^\dagger_g}, \\
  \mathcal{P}_\lambda(\op{\rho}_{SD}) &= \frac{1}{2}\Trd{\op{U}_g (e^{i\lambda \op{p}} \op{\rho}_{SD} + \op{\rho}_{SD}e^{i\lambda \op{p}}) \op{U}^\dagger_g}, 
\end{align}
\end{subequations}

Computing derivatives of the characteristic functions produces the conditioned detector moments,
\begin{subequations}
\begin{align}
  \cmean{f}{x^n} = \frac{\partial^n}{\partial(i\lambda)^n} \cmean{f}{e^{i\lambda x}}\Big|_{\lambda = 0}, \\
  \cmean{f}{p^n} = \frac{\partial^n}{\partial(i\lambda)^n} \cmean{f}{e^{i\lambda p}}\Big|_{\lambda = 0}.
\end{align}
\end{subequations}
This procedure is similar in spirit to the full counting statistics approach employed in \cite{DiLorenzo2012}.

The first two moments are given explicitly by,
\begin{subequations}
\begin{align}
  \cmean{f}{x} &= \text{Re}\mean{x}^w  + g\, \text{Re}\mean{A}^w , \\
  \cmean{f}{p} &= \text{Re}\mean{p}^w, \\
  \cmean{f}{x^2} &= \text{Re}\mean{x^2}^w + 2 g\, \text{Re}\mean{x A}^w + g^2\, \text{Re}\mean{A^2}^w, \\
  \cmean{f}{p^2} &= \text{Re}\mean{p^2}^w,
\end{align}
\end{subequations}
in terms of the Heisenberg evolved joint post-selection $\op{P}_{SD}' = \op{U}_g^\dagger(\op{P}_f\otimes\op{1}_D)\op{U}_g$ and the joint weak values,
\begin{subequations}
\begin{align}
  \mean{x}^w &= \frac{\Tr{\op{P}_{SD}'\, (\op{1}_S\otimes\op{x})\, \op{\rho}_{SD}}}{\Tr{\op{P}_{SD}'\, \op{\rho}_{SD}}}, \\
  \mean{A}^w &= \frac{\Tr{\op{P}_{SD}'\, (\op{A}\otimes\op{1}_D)\, \op{\rho}_{SD}}}{\Tr{\op{P}_{SD}'\, \op{\rho}_{SD}}}, \\
  \mean{p}^w &= \frac{\Tr{\op{P}_{SD}'\, (\op{1}_S\otimes\op{p})\, \op{\rho}_{SD}}}{\Tr{\op{P}_{SD}'\, \op{\rho}_{SD}}}, \displaybreak[0]\\
  \mean{x^2}^w &= \frac{\Tr{\op{P}_{SD}'\, (\op{1}_S\otimes\op{x}^2)\, \op{\rho}_{SD}}}{\Tr{\op{P}_{SD}'\, \op{\rho}_{SD}}}, \displaybreak[0]\\
  \mean{A x}^w &= \frac{\Tr{\op{P}_{SD}'\, (\op{A}\otimes\op{x})\, \op{\rho}_{SD}}}{\Tr{\op{P}_{SD}'\, \op{\rho}_{SD}}}, \displaybreak[0]\\
  \mean{A^2}^w &= \frac{\Tr{\op{P}_{SD}'\, (\op{A}^2\otimes\op{1})\, \op{\rho}_{SD}}}{\Tr{\op{P}_{SD}'\, \op{\rho}_{SD}}}, \\
  \mean{p^2}^w &= \frac{\Tr{\op{P}_{SD}'\, (\op{1}_S\otimes\op{p}^2)\, \op{\rho}_{SD}}}{\Tr{\op{P}_{SD}'\, \op{\rho}_{SD}}}.
\end{align}
\end{subequations}

\subsection*{Detector Wigner Function}
Assuming an initial product state $\op{\rho}_{SD} = \op{\rho}_S\otimes\op{\rho}_D$, we can compute the operations \eqref{eq:opslambda} as follows.  After computing the detector trace in the $p$-basis and inserting two complete $x$-basis sets, the $\mathcal{P}_\lambda$ operation takes the form
\begin{align}
  \mathcal{P}_\lambda(\op{\rho}_{SD}) &= \iiint \frac{dpdxdx'}{2\pi\hbar}\, \bra{x'}\op{\rho}_D\ket{x} \nonumber \\
  &\qquad \qquad e^{-i\frac{p}{\hbar}(x - x' - g\,\text{ad}[\op{A}] + \hbar\lambda)}(\op{\rho}_S), \\
  &= \iint dxdx'\, \bra{x'}\op{\rho}_D\ket{x} \nonumber \\
  &\qquad \qquad \delta(x - x' - g\,\text{ad}[\op{A}] + \hbar\lambda)(\op{\rho}_S), \nonumber\\
  &= \int dz\, \widetilde{W}_D(z, g\,\text{ad}[\op{A}] - \hbar\lambda)(\op{\rho}_S).\nonumber
\end{align}
Here we have changed integration variables to $z = x - x'$ and $y = (x+x')/2$, and have noted that $\widetilde{W}_D(z,y) = \bra{z-y/2}\op{\rho}_D\ket{z+y/2}$ is the Fourier-transformed Wigner function of the detector.  

Performing a similar computation for $\mathcal{X}_\lambda$ yields,
\begin{align}
  \mathcal{X}_\lambda(\op{\rho}_{SD}) &= \iiint \frac{dpdxdx'}{2\pi\hbar}\, \frac{1}{2}(e^{i\lambda x'}+e^{i\lambda x}) \bra{x'}\op{\rho}_D\ket{x} \nonumber\\
  &\qquad \qquad e^{-i\frac{p}{\hbar}(x - x' - g\,\text{ad}[\op{A}])}(\op{\rho}_S), \\
  &= \iint dxdx'\, \frac{1}{2}(e^{i\lambda x'}+e^{i\lambda x}) \bra{x'}\op{\rho}_D\ket{x} \nonumber \\
  &\qquad \qquad \delta(x - x' - g\,\text{ad}[\op{A}])(\op{\rho}_S), \nonumber \\
  &= \int dz\, e^{i\lambda z} \widetilde{W}_D(z,g\,\text{ad}[\op{A}])\cos\left(\frac{\lambda g}{2}\text{ad}[\op{A}]\right)(\op{\rho}_S). \nonumber
\end{align}
Taking derivatives with respect to $(i\lambda)$ produces the expressions in the main text for the first moments.  Setting $\lambda=0$ in either $\mathcal{P}_\lambda(\op{\rho}_S)$ or $\mathcal{X}_\lambda(\op{\rho}_S)$ produces the post-interaction reduced system state $\op{\rho}_S'$.

The operation $\text{ad}[\op{A}]$ is linear, so any analytic function of $\text{ad}[\op{A}]$ may be defined via its Taylor series in the same manner as an analytic function of a matrix.  Indeed, to more rigorously perform the above derivations one can exploit an isomorphism that maps $\op{\rho}_S$ into a vector and $\text{ad}[\op{A}]$ into a matrix acting on that vector.  After expanding the expressions into the eigenbasis of the matrix of $\text{ad}[\op{A}]$ and regularizing any singular functions into limits of well-behaved analytic functions, the above computations can be performed for each eigenvalue, summed back into a matrix, and then mapped back into the operator form shown.  For unbounded $\op{A}$ then one must also carefully track the domains to ensure that the resulting expressions properly converge, as they should for any physically sensible result.

\subsection*{Hermite-Gauss superpositions}
The Wigner distribution for an arbitrary superposition of Hermite-Gauss modes $\ket{\psi} = \sum_m c_m \ket{h_m}$ can be computed to find (suppressing arguments for compactness),
\begin{align}\label{eq:hgwignersup}
  W &= \sum_{m,n=0}^\infty \frac{c_m c_n^*}{\sqrt{m!\, n!}}\frac{(-1)^m\,e^{i(m-n)\phi} }{\pi \hbar} D^m_n[\sqrt{2G}] \,e^{-G},
\end{align}
where $G(x,p) = x^2/2\sigma^2 + 2p^2\sigma^2/\hbar^2$ and $\phi(x,p) = \tan^{-1}(-2p\sigma^2 / \hbar x)$.  The polynomial sequence $D^m_n(x)$ has the generating function,
\begin{align}\label{eq:hgpolygen}
  \exp(z \bar{z} - x(z - \bar{z})) = \sum_{m,n=0}^\infty \frac{z^m \bar{z}^n}{m!\, n!} D^m_n(x),
\end{align}
and the explicit form,
\begin{align}\label{eq:hgpoly}
  D^m_n(x) &= \sum_{k=0}^{\text{min}(m,n)} \frac{m!\, n!\,(-1)^{m-k}}{(m-k)!\,(n-k)!\,k!}\, x^{m+n - 2k}.
\end{align}
Notably, the diagonal elements of this sequence are the Laguerre polynomials, $D^m_m(x) = m!\,L_m(x^2)$.  These results can be obtained by using the generating function for the Hermite polynomials $\exp(2xz - z^2) = \sum_{n=0}^\infty H_n(x) z^n / n!$, as well as the identities $H_n(x) = \exp(-\partial_{2x}^2) (2x)^n$ and $\int dx\, e^{-x^2} H_n(x) H_m(x) = \delta_{m,n}\, m!\, \sqrt{\pi}\, 2^m$.

Computing the reduced system state $\op{\rho}_S'$ using this Wigner function yields,
\begin{align}\label{eq:sysredhgsup}
  \op{\rho}_S' &= \sum_{m,n=0}^\infty \frac{c_m c_n^*}{\sqrt{m!\, n!}}\, D^m_n\left[\sqrt{\epsilon}\,\text{ad}[\op{A}]\right]\, e^{\epsilon \mathcal{L}[\op{A}]}(\op{\rho}_S),
\end{align}
where $\epsilon = (g/2\sigma)^2$, and $\mathcal{L}[\op{A}] = -\text{ad}^2[\op{A}]/2$ is the Lindblad decoherence operation.  Notably, the functional form of the Wigner distribution \eqref{eq:hgwignersup} is still largely preserved in Eq.~\eqref{eq:sysredhgsup}.  

The detector averages can also be computed from this Wigner function.  The weak value $\text{Re}\mean{x}^w$ will vanish by symmetry; the weak value $\text{Re}\mean{p}^w$ involves the derivative $i\hbar\partial_{g\text{ad}[\op{A}]} \op{\rho}_S'$; and, the weak value $\text{Re}\mean{A}^w$ involves the state $\op{\rho}_S'$ directly.  When $c_m = 1$ with the rest of the coefficients zero, then these generalizations reduce to the results presented in the main text.

\end{document}